\documentclass[10pt]{iopart}

\usepackage{iopams}
\usepackage{hyperref}
\hypersetup{colorlinks=true, urlcolor=red, citecolor=blue, linkcolor=blue}
\usepackage{graphicx}
\usepackage{subfigure}
\usepackage{epstopdf}
\usepackage{color}
\usepackage{amssymb}
\usepackage{amsfonts}
\usepackage{bm}
\usepackage{epsfig}
\usepackage{cite}
\usepackage{latexsym}

\expandafter\let\csname equation*\endcsname\relax

\expandafter\let\csname endequation*\endcsname\relax

\usepackage{amsmath}
\usepackage{color}

\usepackage{ulem}
\usepackage{soul}

\newcommand{\lrang}[1]{\langle #1\rangle}

\begin{document}

\renewcommand*{\DefineNamedColor}[4]{%
   \textcolor[named]{#2}{\rule{7mm}{7mm}}\quad
  \texttt{#2}\strut\\}

\definecolor{red}{rgb}{1,0,0}

\title[Constructive interference between disorders for entanglement in spin models]{Constructive Interference Between Disordered Couplings Enhances Multiparty Entanglement in Quantum
Heisenberg Spin Glass Models}

\author{Utkarsh Mishra, Debraj Rakshit, R. Prabhu, Aditi Sen(De) and Ujjwal Sen}

\address{
Harish-Chandra Research Institute, Chhatnag Road, Jhunsi, Allahabad 211 019, India }

\begin{abstract}
Disordered systems form one of the centrestages of research in many body sciences and lead to a 
plethora of interesting phenomena and applications. A paradigmatic disordered system consists of 
an one-dimensional array of quantum spin-1/2 particles, governed by the Heisenberg spin glass 
Hamiltonian with natural or engineered quenched disordered couplings in an external magnetic 
field. These systems allow disorder-induced enhancement
for bipartite and multipartite observables.
Here we show that simultaneous application of independent quenched 
disorders results in disorder-induced enhancement, while the same is absent with individual application of 
the same disorders. We term the phenomenon as constructive interference and the corresponding parameter stretches as the Venus regions. Interestingly, it has only been observed for multiparty entanglement and is absent for the single- and 
two-party physical quantities.
\end{abstract}

\maketitle

\section{Introduction}

 Research in implementation of quantum devices has led to the identification of useful multiparty quantum  information processing tasks, like quantum secret sharing \cite{secret-sharing}, cluster state quantum computation \cite{cluster-state-comp1,cluster-state-comp2,cluster-state-comp3}, quantum state transmission \cite{quantum-state-trans}, and  distributed quantum dense coding \cite{quantum-dense-coding-1,quantum-dense-coding-2,quantum-dense-coding-3}. These multisite tasks have already been 
implemented in several physical systems like ion-traps \cite{ion-trap-1,ion-trap-2}, photons \cite{photon}, optical lattices \cite{opt-lattices}, and superconducting qubits \cite{supercqubit-1,supercqubit-2,supercqubit-3}. It is widely believed that understanding the role of bipartite and multipartite entanglement \cite{bi-multy-quantum-info-process} in these protocols in particular and quantum many-body systems in general, is crucial to gain the ability to build scalable decoherence-resistant efficient quantum devices.
The wide interest in such  activities is also due to the fact that several concepts developed in quantum information science turn out to be useful tools to detect co-operative phenomena \cite{amader-adp,fazioRMP, Sachdev}, like quantum phase transitions and thermal transitions, and can help to develop approximate methods to obtain the ground states of non-integrable  systems \cite{appoximategroundstate}.

 The importance of studying the effect of disorder in many-body systems can hardly be overestimated \cite{realsystem-impurities,eibar-eibar-khukumoni-uthbe-1,eibar-eibar-khukumoni-uthbe-2,eibar-eibar-khukumoni-uthbe-3,eibar-eibar-khukumoni-uthbe-4,eibar-eibar-khukumoni-uthbe-5,eibar-eibar-khukumoni-uthbe-6,eibar-eibar-khukumoni-uthbe-7,Igloi2005,Misguich2005}. Realization of most physical systems inherently  results in impurities or defects, which may suppress the physical properties of the systems \cite{proplossdisorder-1,proplossdisorder-2,proplossdisorder-3,proplossdisorder-4,proplossdisorder-5}. However, disordered systems, both  classical and quantum, display  counterintuitive phenomena like disorder-induced enhancement in several physical quantities like magnetization, classical correlators, and entanglement (see \cite{Misguich2005,order-from-dis-1,order-from-dis-2,order-from-dis-3,order-from-dis-4,qc-more-in-disorder-1,qc-more-in-disorder-2} and references therein). At the same time, disordered systems sustain rich phases like spin glass \cite{spinglass-1,spinglass-2,spinglass-3,spinglass-4} and  Bose glass \cite{Boseglass-1,Boseglass-2}, and phenomena like Anderson localization \cite{Alocal} and  high $\textrm{T}_{\textrm{c}}$-superconductivity \cite{hightTsupc-1,hightTsupc-2,hightTsupc-3}. Recent experimental developments, especially in ultra-cold gases, give rise to the possibility of introducing disorder in a controlled way 
and hence paves the way for novel recipes of observation of these properties in the laboratory. 

 In this paper, we concentrate on the behavior of different physical quantities for the ground state of one-dimensional quenched disordered  quantum Heisenberg (or $XYZ$) models or quantum Heisenberg spin glass models. Specifically, we consider three paradigmatic classes of disordered Heisenberg spin glass Hamiltonians: the quenched disorder is in  (a) the ``planar'' couplings, (b) the ``azimuthal'' couplings, and in (c) both the planar and azimuthal couplings. The main results of the paper are as follows: (i)  We find that both bipartite and multipartite  entanglement can be enhanced, in some parameter space of the Hamiltonian, by the introduction of \textit{all} the disorder combinations, mentioned in (a),  (b) and (c).
(ii) We find that in all these models, there are large surfaces in the parameter space in which the magnetization and classical correlators behave in a complementary way to bipartite and multipartite entanglement.
(iii)  More important, and rather engrossing is the uncovering  of parameter ranges where the \textit{individual} insertions of planar and azimuthal quenched disorder couplings do not result in disorder-induced enhancement of a multiparty entanglement measure, while the same appears in the \textit{simultaneous} presence of the disorders. We term the phenomenon as constructive interference of the disordered couplings and call the coupling parameter ranges as the Venus regions.
(iv) Importantly, such constructive interference is not observed in  single- as well as two-site physical quantities like magnetization, classical correlators, and bipartite entanglement. Moreover, changing the Hamiltonian, for example, to the $XY$ model also wipes out the phenomenon.

 The counterintuitive nature of constructive interference for a physical quantity leads us to believe that it can have implications in fundamental and applicational regimes. Moreover, multiparty quantum information processing tasks typically have origins in the bipartite domain. Instances where the converse occurs are few and far between, and indicates important diversions from the usual track (see e.g. \cite{cluster-state-comp1,cluster-state-comp2,cluster-state-comp3 ,CSS-1,CSS-2,threequbit,boundBell-1,boundBell-2,boundBell-3,boundBell-4,boundBell-5,boundBell-6,boundBell-7}). The fact that constructive interference is observed only for multipartite entanglement in the presence of impurities is also in the spirit of these latter instances.

\section {The Heisenberg quantum spin glasses and enhancement score}
We introduce the four different Heisenberg Hamiltonians which are studied in the paper.

\noindent{\texttt{\textbf{Case 0}:}}  The one-dimensional  disordered quantum Heisenberg (or $XYZ$) model with nearest-neighbour interactions  in an external magnetic field is described by the Hamiltonian 
\begin{gather}
H_{\lrang{J,\delta}}=\kappa \left[\sum_{\langle { i,j}\rangle} \frac{J_{ij}}{4} \left[(1+\gamma)\sigma_i^x\sigma_{j}^x+(1-\gamma)\sigma_i^y\sigma_{j}^y\right]\right.
 \left. + \sum_{\langle {i,j}\rangle}\frac{\delta_{ij}}{4} \sigma_i^z\sigma_{j}^z
-\frac{h}{2}\sum_{i}^N \sigma_i^z \right].
\label{eq:HeisenbergHam}
\end{gather}
Here, $J_{ij}(1-\gamma)$ and $J_{ij}(1+\gamma)$ are proportional to the $xx$ and $yy$ interactions, while $\delta_{ij}$ is that to the $zz$ one. \(N\) is the number of spins. $\gamma$ measures the anisotropy between the first two interactions, and is dimensionless. $J_{ij}$, $\delta_{ij}$, and $h$ are also dimensionless. $\kappa$ is a constant, and has the units of energy. $J_{ij}$ are independently and identically distributed (i.i.d.) Gaussian random variables with mean $\langle J \rangle$ and unit standard deviation. Similarly, $\delta_{ij}$ are i.i.d. Gaussian random variables with mean $\langle \delta \rangle$ and unit standard deviation. We set $\lrang{\lambda}=\lrang{J}/h$ and $\lrang{\mu}=\lrang{\delta}/h$, which are therefore again dimensionless.  $\sigma_i^{k}\, (k=x,y,z)$ are the Pauli spin matrices at the $i^{\textrm{th}}$ site and $\lrang{ij}$ indicates that the corresponding summation is over nearest-neighbour spins. The applied field, $h$, is kept ordered throughout the paper. 
{\color {black}In this work, we consider periodic and open boundary conditions for  systems with $N<10$ and $N>10$, respectively. For the relatively larger systems with open boundary condition, we calculate the local observables (the one- and two-site quantities) at the center of the chain. For example, we investigate magnetization of the $(N/2)^{th}$ site, and the two-site observables, such as correlators and bipartite entanglement, corresponding to the ($N/2, N/2+1$) pair.}

\noindent{\texttt{\textbf{Case 1}:}}  Quantum Heisenberg  model. In this case, the Hamiltonian, which we denote by $H$,  has site-independent couplings, i.e.,  $J_{ij}=J$ and $\delta_{ij}=\delta$. Since we will be in need of multisite state characteristics, the Bethe ansatz \cite{Bethe}  is difficult to apply in an efficient way, especially in the disordered cases considered.
We denote $J/h$ and $\delta/h$ as $\lambda$ and $\mu$ respectively.

\noindent{\texttt{\textbf{Case 2}:}} Quantum Heisenberg ``planar'' spin glass. In this case, the planar couplings, $J_{ij}$'s are disordered and chosen from i.i.d Gaussian distribution with mean $\lrang{\lambda}=\lrang{J}/h$ and unit standard deviation, while the couplings $\delta_{ij}$ are considered to be site-independent, and fixed at $\delta$. In analogy with Eq.~(\ref{eq:HeisenbergHam}), we denote the Hamiltonian by $H_{\lrang{J}}$. 


\noindent{\texttt{\textbf{Case 3}:}} Quantum Heisenberg ``azimuthal'' spin glass. The system in this case is governed by the Hamiltonian, $H_{\lrang{\delta}}$, in which $J_{ij}=J$,   while the couplings, $\delta_{ij}$, are i.i.d. Gaussian random variables with mean $\lrang{\delta}$ and unit standard deviation.

 In the models that we consider here, the disordered parameters are quenched (see  \ref{appendix_Quenched_averaging}). In a disordered system, if a quenched averaged physical quantity, ${\cal Q}_{av}$, associated with a state of the system is larger than the same quantity, ${\cal Q}$, of the corresponding ordered system in the analogous state, then the value of the physical quantity is said to  exhibit a disorder-induced enhancement.  To characterize such advantage, we introduce the enhancement score, $\Delta^{{\cal Q}}$,  of a physical quantity ${\cal Q}$.
The enhancement scores for the different physical quantities indicate the usefulness of the disordered system in comparison with the corresponding ordered system with respect to that physical quantity. It is a quantification of the usefulness, and is exactly the amount of the physical quantity in the disordered system which exceeds that in the ordered system. 
  More precisely, we set
\begin{equation}
\label{eq:disorderscore}
  \Delta^{\cal Q}_{a,b,\ldots}=|{\cal Q}_{av}(\langle a \rangle,\langle b \rangle,\ldots)|-|{\cal Q}(\lrang{a},\lrang{b}, \ldots)|.
\end{equation}
Here, ${\cal Q}_{av}(\langle a \rangle,\langle b \rangle,\ldots)$ is the quenched averaged value of a physical quantity, ${\cal Q}$, of the system where the averaging is performed over the system parameters, $a,\, b,\ldots$, which follow Gaussian distributions with mean $\lrang{a}$, $\lrang{b},\ldots$ and standard deviations $\sigma_{a}$, $\sigma_{b},\ldots$ respectively. The definition can of course be generalized to the case of other probability distributions. ${\cal Q}(\lrang{a},\lrang{b},\ldots)$ is the corresponding physical quantity for the ordered case of the same system, where the values of the system parameters $a,b,\ldots$  are kept constant (i.e., they are not disordered) at $\lrang{a},\lrang{b},\ldots$  respectively. Both ${\cal Q}_{av}$ and ${\cal Q}$  usually depend also on other system parameters (that are not disordered) which are kept the same for both the systems (disordered and ordered) and which are kept silent in the notation. A positive enhancement score for a physical quantity, ${\cal Q}$, in a certain range of the system parameters will imply that ``disorder-induced enhancement'' or ``disorder-induced enhanement'' is attained for ${\cal Q}$ in that region of the parameter space. Whereas, a negative value of the same will indicate that ${\cal Q}$ gets degraded.

We now investigate the behavior of different measurable quantities in all the three disordered models viz Cases 0, 2 and 3. We compute the ground state in each of these models and investigate the behavior of the enhancement score, $\Delta^{\cal Q}_{\xi}$, corresponding to physical quantities like, magnetization, $M_z$, which is a single-site observable, nearest-neighbor classical correlator, $T_{zz}$, and concurence, $C$, which are two-site observables, and genuine geometric measure, $\cal{E}$ as a multipartite quantity (see \ref{appendix_quantities} for the definitions of these quantities). Here, $\xi$ denotes the aggregate of system parameters that are quenched disordered. There is a wide range of the anisotropy parameter, $\gamma$, and magnetic field strength, $h$, of these models for which disorder-induced enhancement phenomena is observed. Also for a range of values of these parameters the complementarity behavior between classical and quantum quantities are observed.  In general, these features of all the observables remain qualitatively similar with the variation of $\gamma$ and $h$. However, we observe that the phenomena of constructive interference is observed beyond a certain value of the critical magnetic field, $h_{c}$, which depends on the system size. If the system size is five or more than five spins, $h_{c}$ is approximately 0.7.  For the purpose of depiction of  the effects, throughout this paper, we choose $\gamma=0.7$ and $h=0.8$.
However, the effects are 
In the following, we begin our discussion by reporting disorder-induced enhancement for the quantities mentioned above, where disorder corresponds to the Cases 0, 2, and 3 respectively. We highlight the complementarity behavior exhibited by the classical and quantum observables in these cases. We then move to discuss the constructive interference phenomena.

While investigating the disordered cases, quenched averaging is performed over $5\times 10^{3}$ realizations. For fixed values of $\lrang{\lambda}, \lrang{\delta}$, and $h$,  we have  performed the numerical simulations for higher number of realizations and have found that the corresponding quenched physical quantities have already converged for $5\times 10^{3}$ realizations or before.

\begin{figure}
\includegraphics[angle=270,width=7cm]{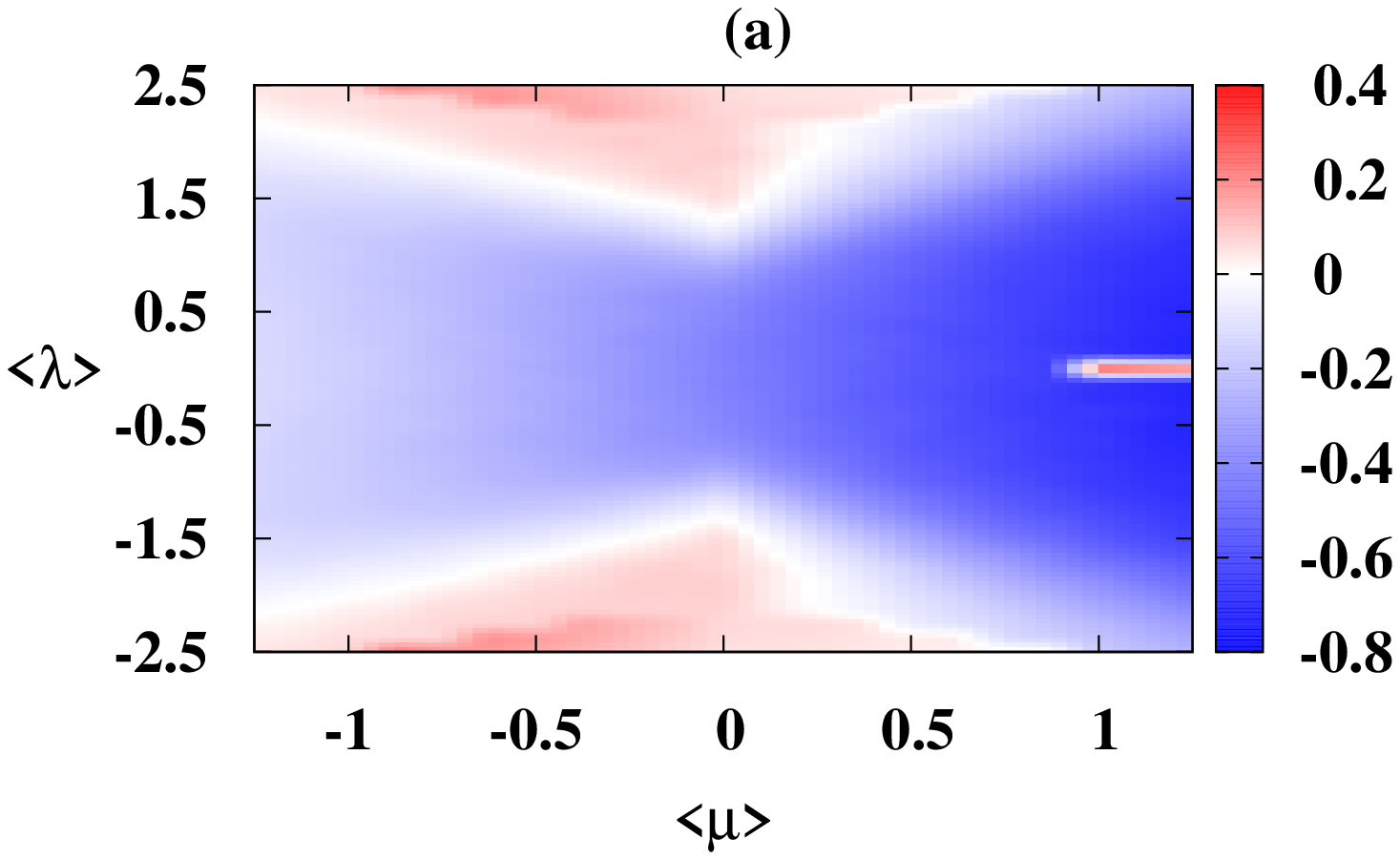}%
\includegraphics[angle=270,width=7cm]{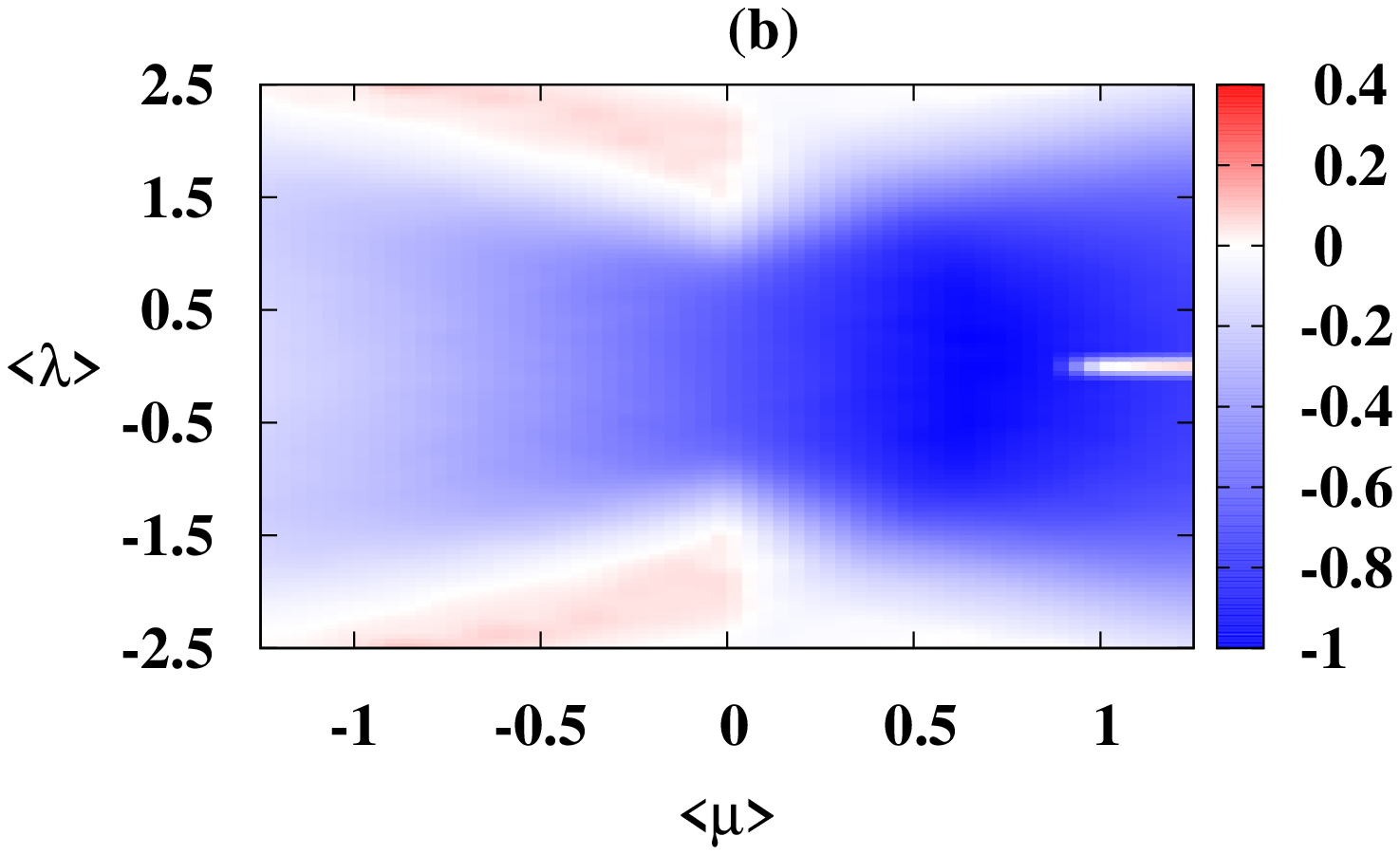}
\includegraphics[angle=270,width=7cm]{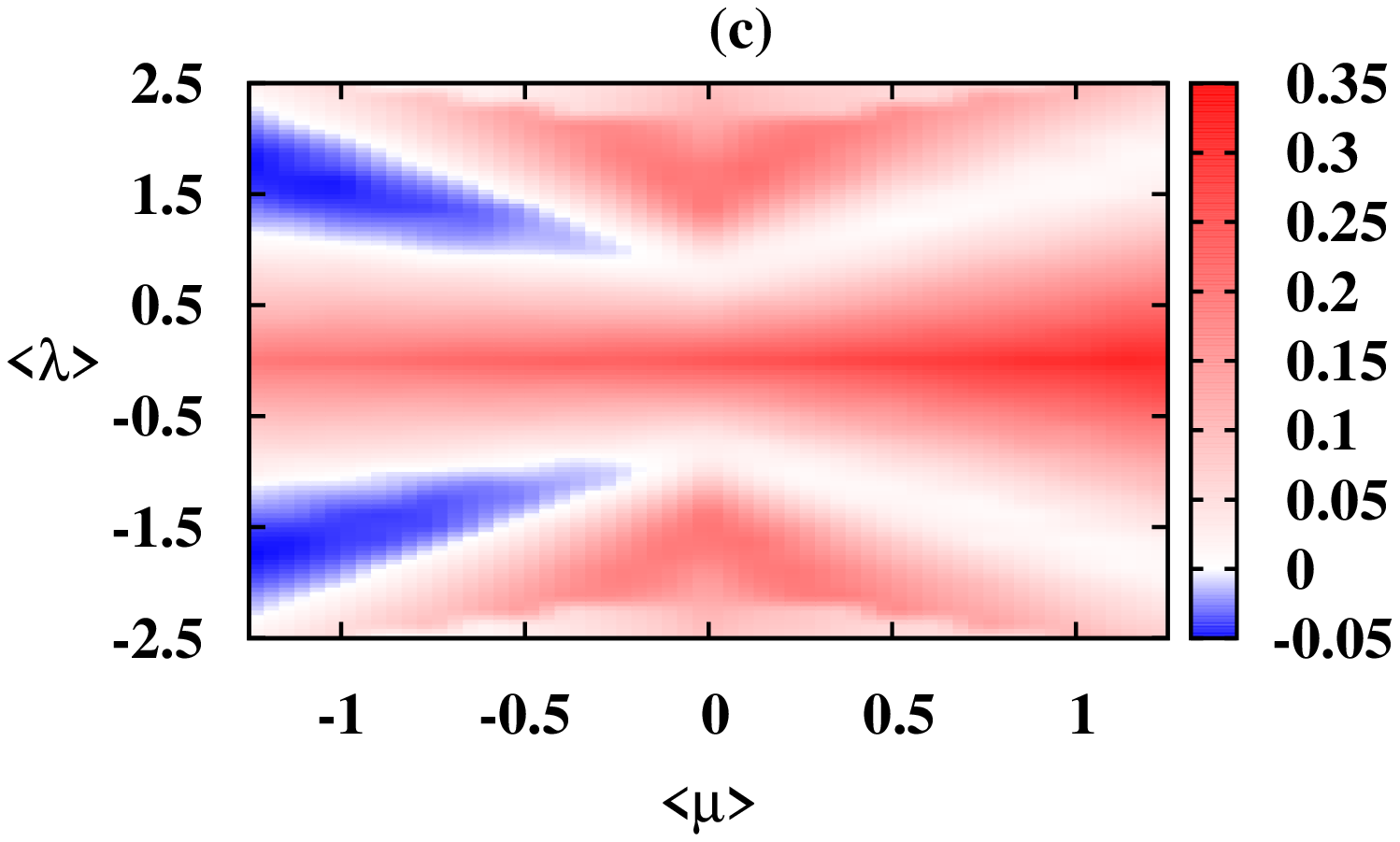}%
\includegraphics[angle=270,width=7cm]{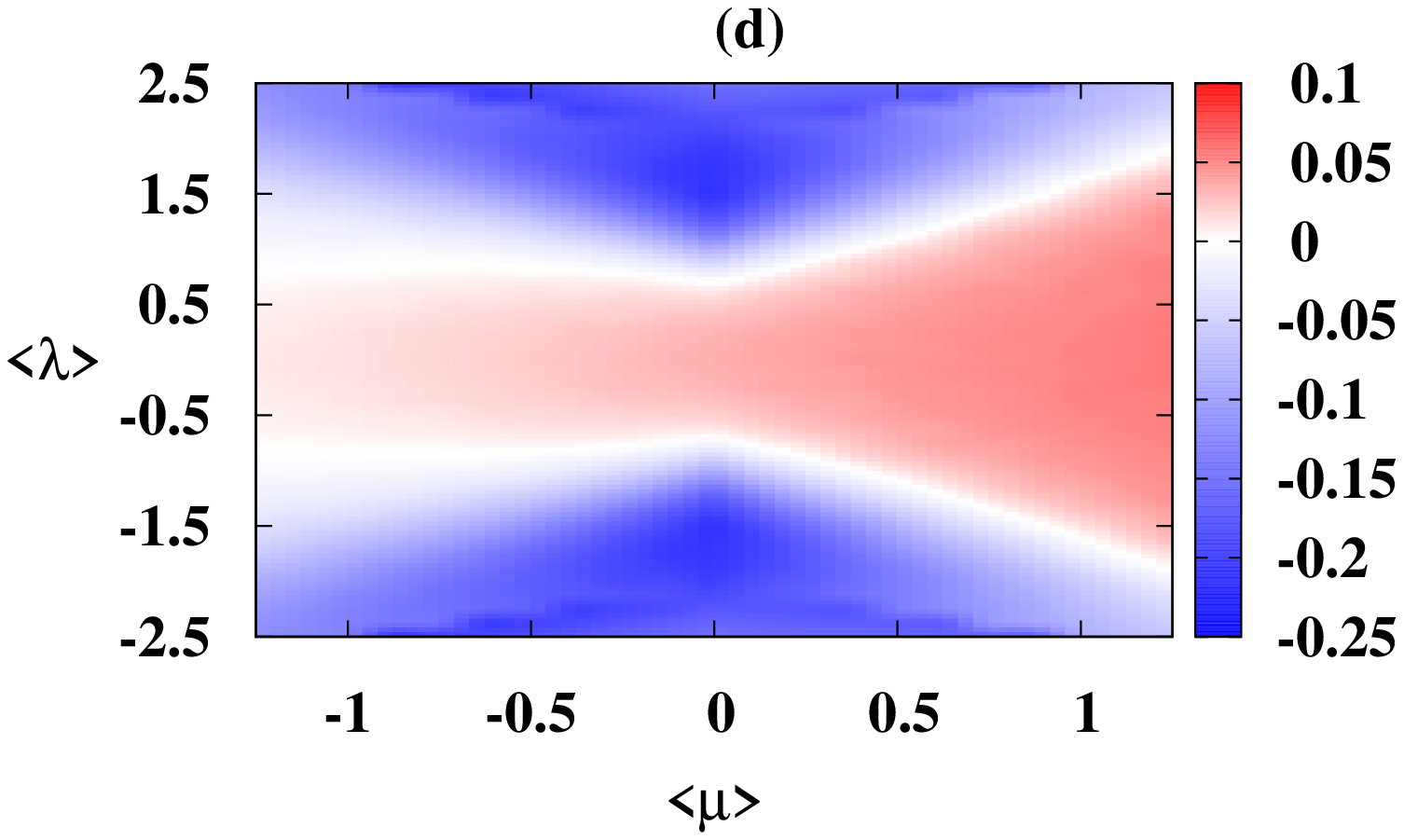}
\caption{Disorder-induced enhancement and complementarity when both planar and azimuthal couplings are quenched disordered. The enhancement scores $(\Delta^{\cal Q}_{\lambda,\mu})$  for (a) magnetization ($\Delta ^{M_{z}}_{\lambda,\mu}$), (b) the $zz$-classical correlator ($\Delta ^{T_{zz}}_{\lambda,\mu}$), (c) bipartite quantum correlation as quantified by concurrence ($\Delta ^{C}_\lambda$), and (d) genuine multipartite quantum correlation measure quantified by generalized geometric measure ($\Delta ^{\mathcal{E}}_{\lambda,\mu}$) for $N=6$.
 In these panels, the quantities along ordinates are $\lrang{\lambda}$, while the abscissae represent the $\lrang{\mu}$. Quenched average of the observable is performed over $5\times 10^{3}$ random realizations. The regions represented in red are the ones for which $\Delta^{\cal Q}$ is positive indicating that the corresponding physical quantity ${\cal Q}$ attains a higher value with the introduction of disorder in these regions. The areas represented in blue are the ones for which $\Delta^{\cal Q}$ is negative and they point to parameter regions where the ${\cal Q}$ is higher in the corresponding clean system. In the white regions, ${\cal Q}$ remains unaltered by the introduction of disorder in the system. All the parameters plotted here are dimensionless. 
}
\label{fig:N6dxydzzcmztzzggm}
\end{figure}

\begin{figure}
\includegraphics[angle=270,width=7cm]{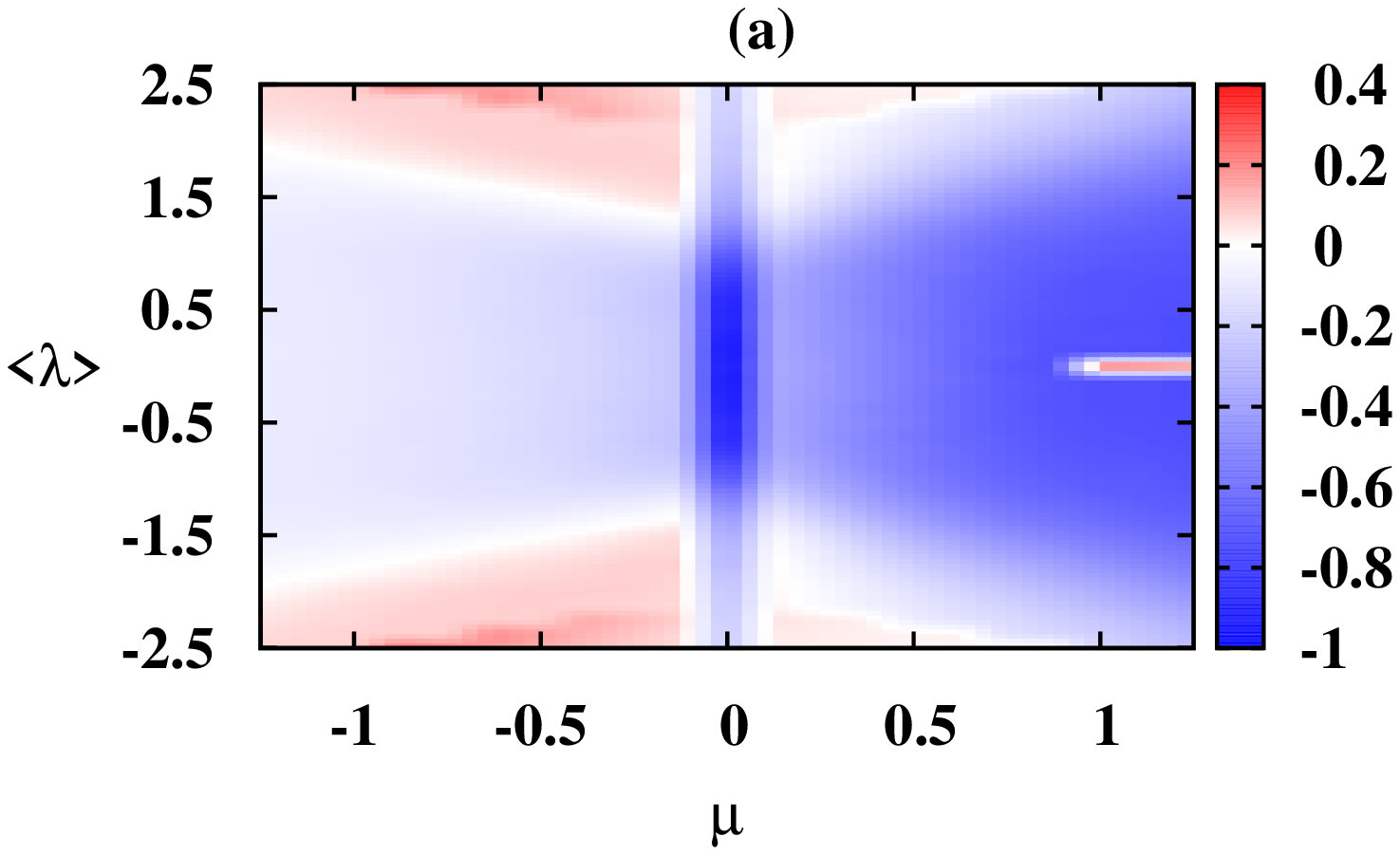}%
\includegraphics[angle=270,width=7cm]{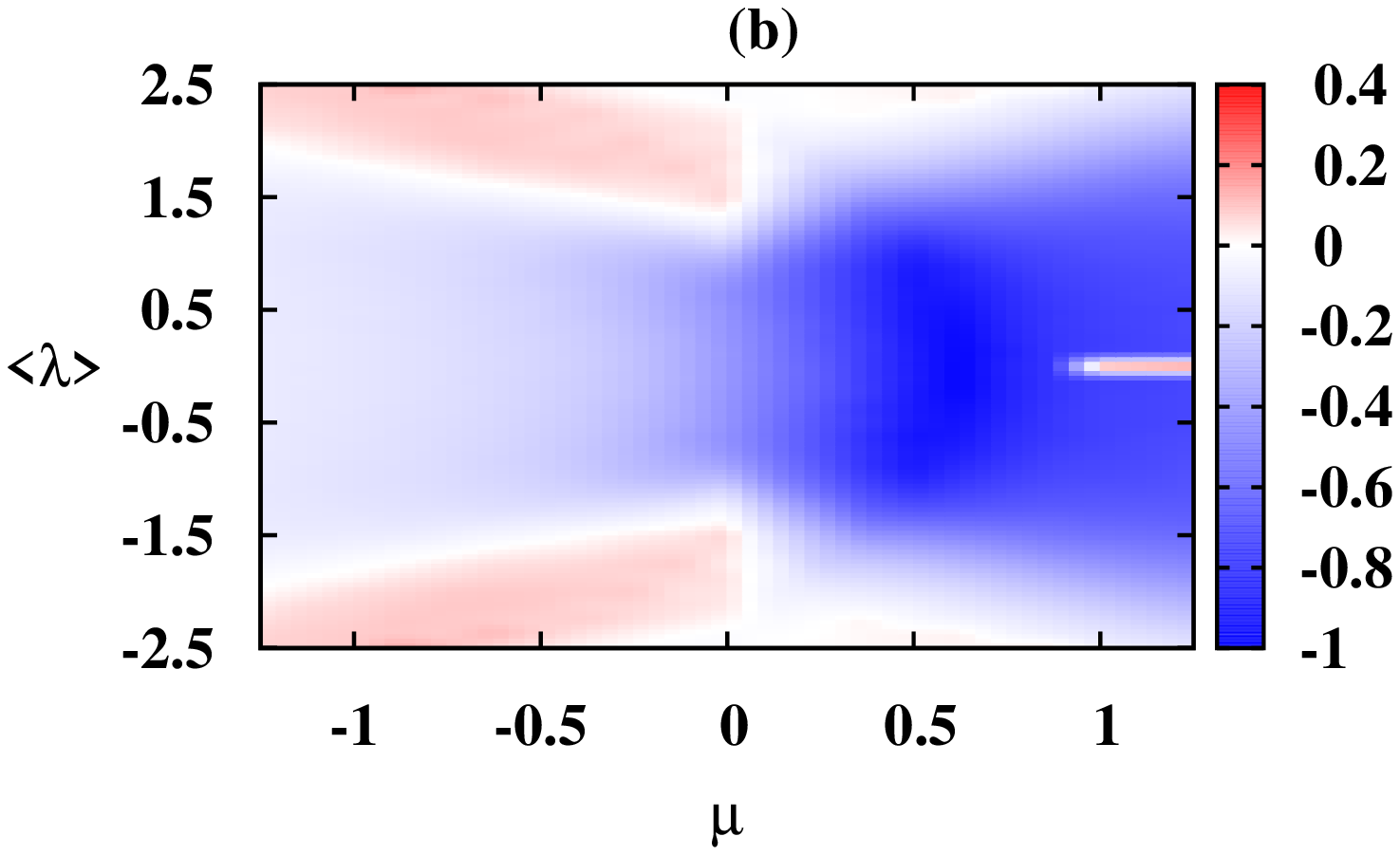}\\
\includegraphics[angle=270,width=7cm]{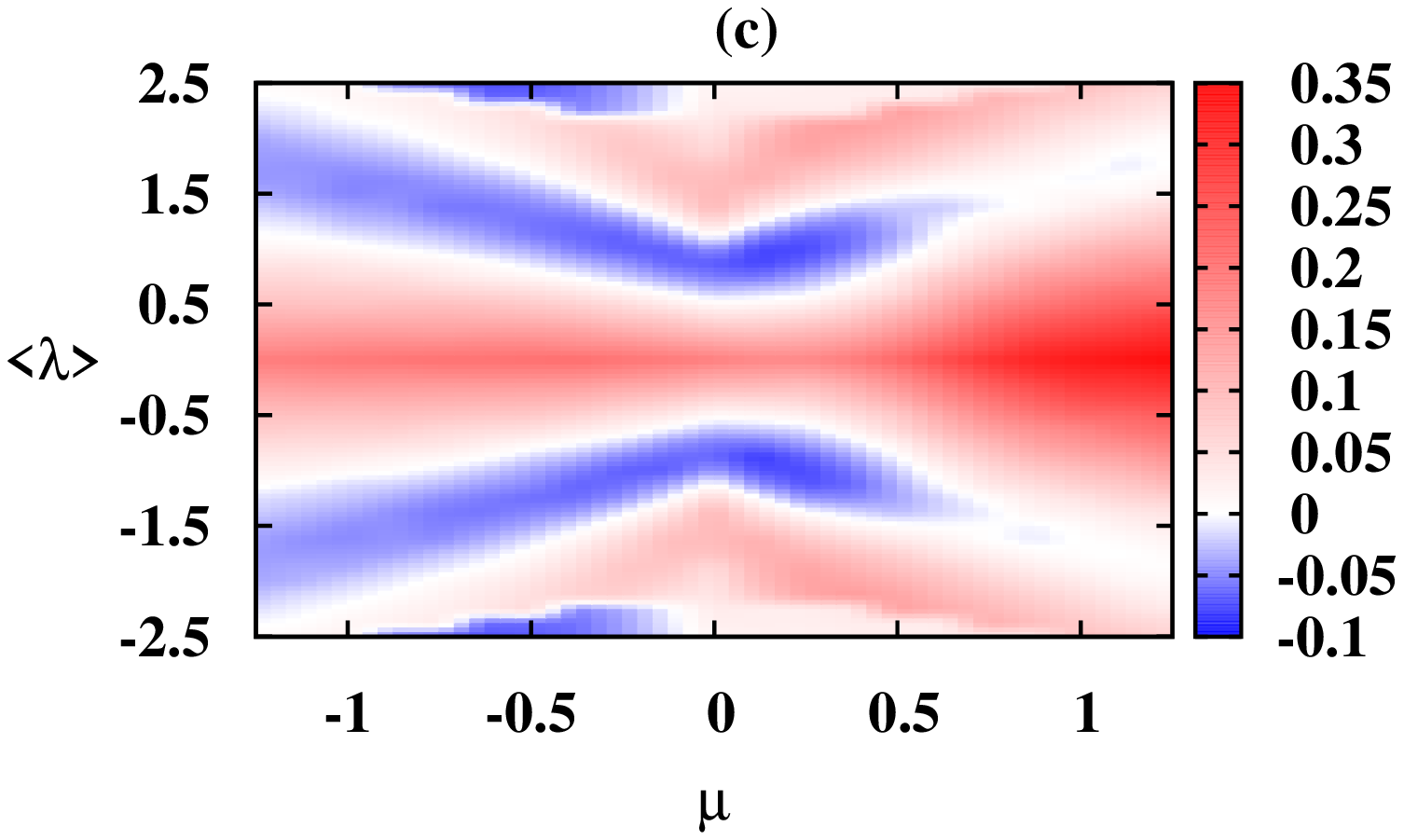}%
\includegraphics[angle=270,width=7cm]{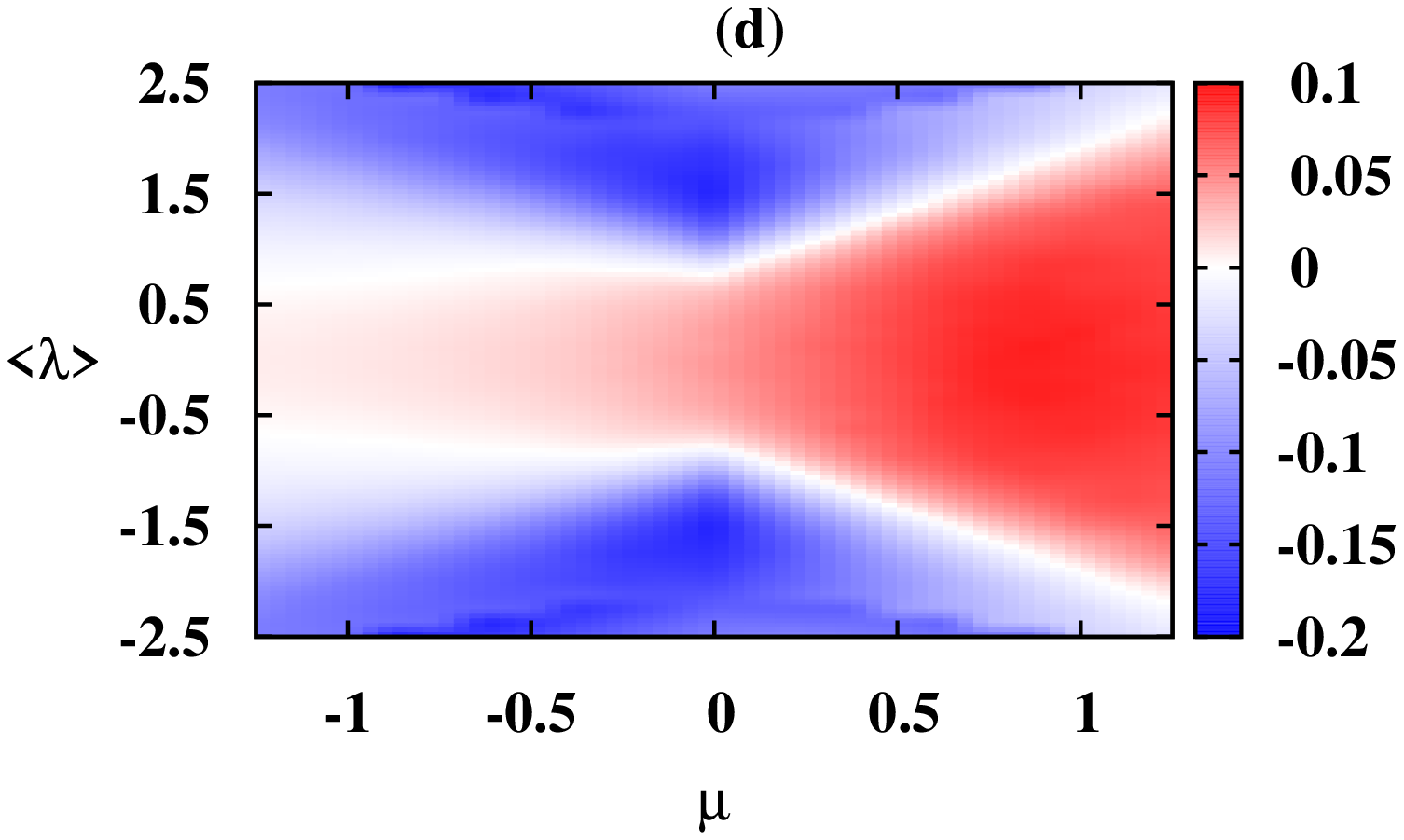}
\caption{Disorder-induced enhancement and complementarity in planar spin glass. The plots in the different panels are for (a) $\Delta ^{M_{z}}_\lambda$, (b) $\Delta ^{T_{zz}}_\lambda$, (c) $\Delta^{C}_\lambda$, and (d) $\Delta ^{\mathcal{E}}_\lambda$. The disordered Hamiltonian for the enhancement scores is $H_{\lrang{\lambda}}$, whose 6 i.i.d. Gaussian random variables, $\lambda_{ij}$, have mean $\lrang{\lambda}$ and unit standard deviation. In each of these panels, the ordinate is $\lrang{\lambda}$ and the abscissa is $\mu$. All other considerations are the same as in Fig.~\ref{fig:N6dxydzzcmztzzggm}.
}
\label{fig:N6dxyozzcmztzzggm}
\end{figure}

\begin{figure}
\includegraphics[angle=270,width=7cm]{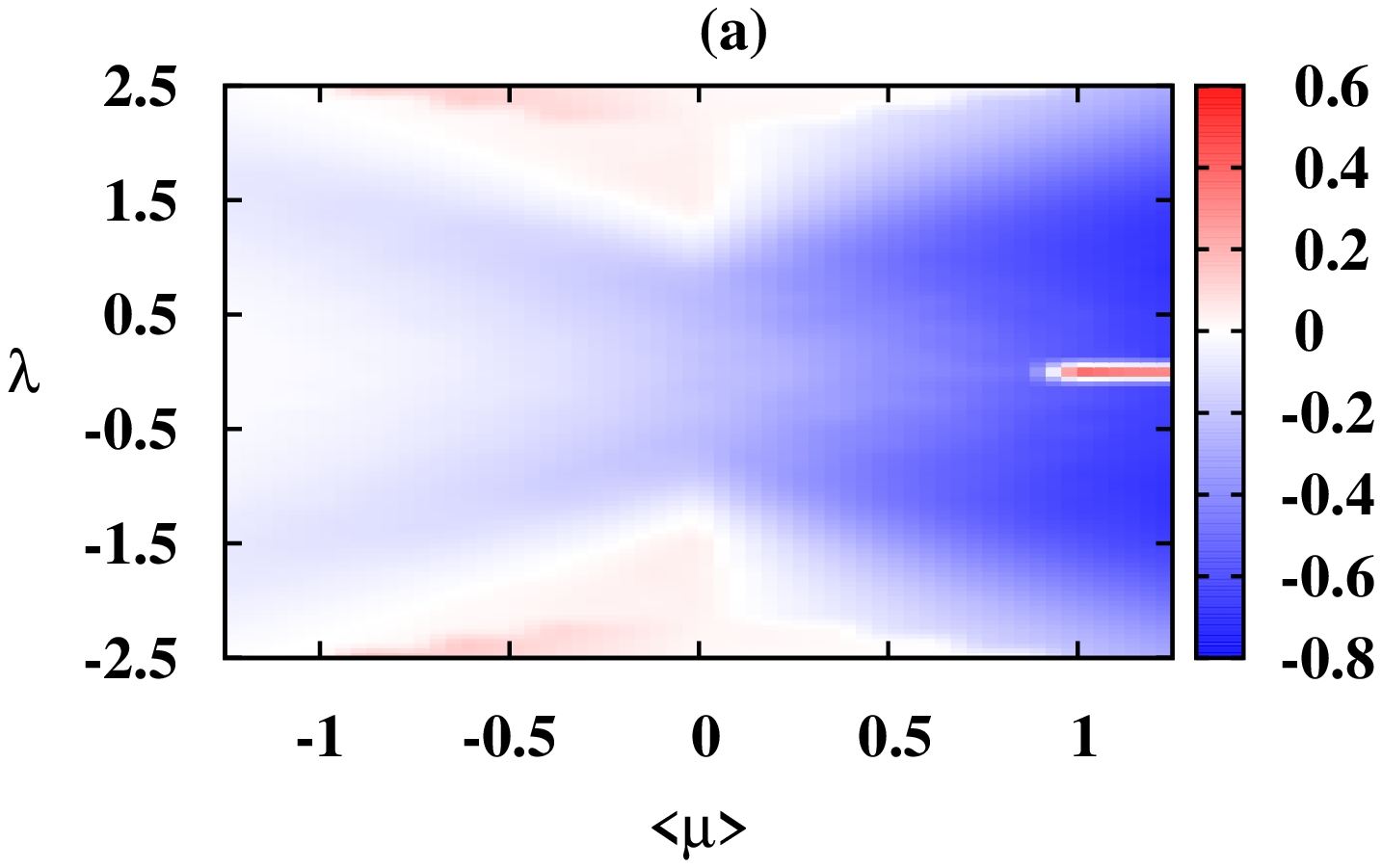}%
\includegraphics[angle=270,width=7cm]{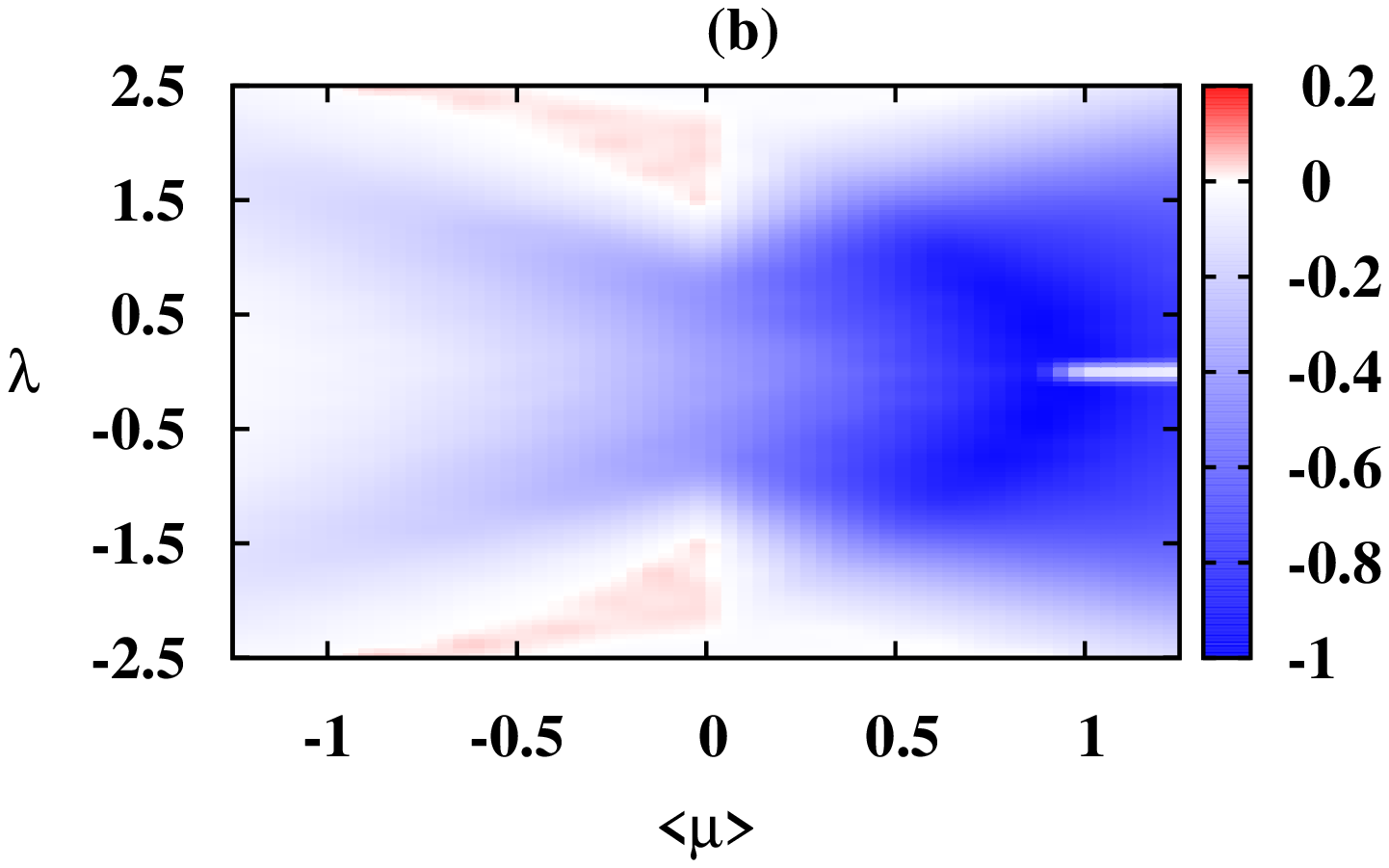}\\
\includegraphics[angle=270,width=7cm]{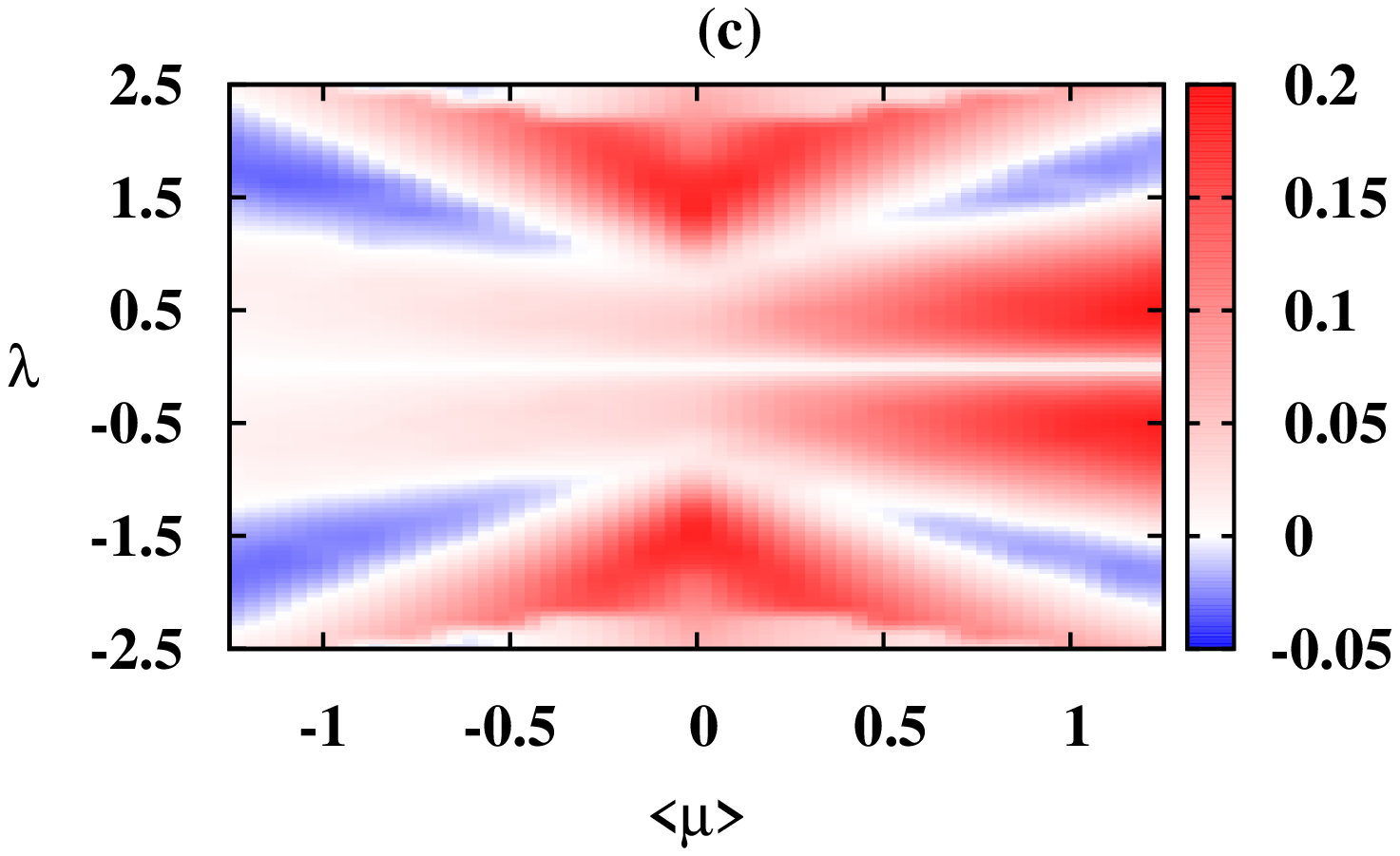}%
\includegraphics[angle=270,width=7cm]{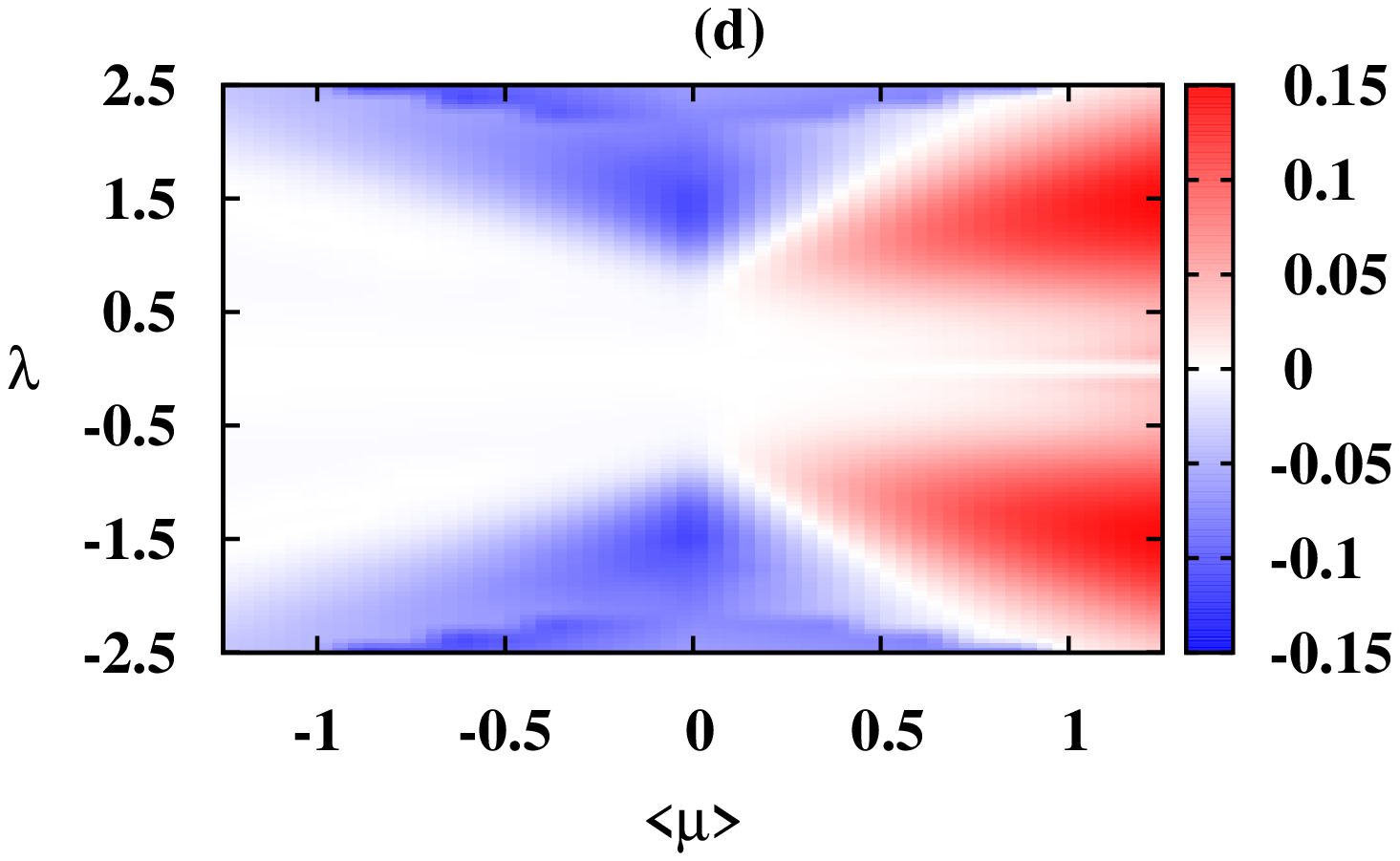}
\caption{Disorder-induced enhancement and complementarity in azimuthal spin glass. The plots in the different panels are for (a) $\Delta ^{M_{z}}_\mu$, (b) $\Delta ^{T_{zz}}_\mu$, (c) $\Delta^{C}_\mu$, and (d) $\Delta ^{\mathcal{E}}_\mu$. The disordered Hamiltonian for the enhancement scores is $H_{\lrang{\delta}}$, whose 6 i.i.d. Gaussian random variables, $\delta_{ij}$, have mean $\lrang{\delta}$ and unit standard deviation. In each of these panels, the ordinate is $\lambda$ and the abscissa is $\lrang{\mu}$. All other considerations are the same as in Fig.~\ref{fig:N6dxydzzcmztzzggm}.
}
\label{fig:N6oxydzzcmztzzggm}
\end{figure}

\section{Disorder-induced enhancement and complementarity in classical and quantum quantities}
We investigate here the behaviour of enhancement scores of  different physical quantities  for the case when the disorder is introduced in both planar and azimuthal couplings  (see Eq. (\ref{eq:HeisenbergHam})). In Fig.~\ref{fig:N6dxydzzcmztzzggm}, we show the behavior of the enhancement scores of magnetization, $zz$-correlator, concurrence, and generalized geometric measure (GGM), with the variation of $\lrang{\lambda}$ and $\lrang{\mu}$. In all the cases considered, we have  observed disorder-induced enhancement also for $T_{xx}$ and $T_{yy}$. We do not exhibit them in the figures given. The investigation shows that for all observables, there exist regions in which the enhancement score is vanishing. These appear as white regions in the panels in the figure. Also, for large mean values of the disordered interactions, where the simultaneous presence of ferro- and anti-ferromagnetic couplings due to the disorder is absent, the enhancement scores vanish. In all the observables considered, viz. magnetization, classical correlators, and bipartite as well as multipartite entanglement, for a given $\lrang{\mu}$, there typically appears oscillations in the surface of the enhancement score as we scan the $\lrang{\lambda}$ axis and occasionally such oscillations have a positive enhancement score in their crests and negative one in their troughs. The parameter regions, which have a positive enhancement for a certain physical quantity, indicates an disorder-induced enhancement for that quantity. See Fig.~\ref{fig:N6dxydzzcmztzzggm} for a depiction. It can also be seen from Fig.~\ref{fig:N6dxydzzcmztzzggm} that there exists wide stretches of parameter regime, where the bi-, and multi-party quantum correlations enhance compared to the clean system even though the classical correlator and the magnetization diminish and vice-versa.  Such complementary features are true for all types of disorder considered here including Case 2 and 3 (See Fig.~\ref{fig:N6dxyozzcmztzzggm} and \ref{fig:N6oxydzzcmztzzggm} respectively).  However, as it is expected, there is a change in parametric regime at which the disorder-induced enhancement appears for different cases of disorder. For example, the behavior of the enhancement scores for magnetization and classical correlators are quite similar for Cases 2 and 3 while this is not true for concurrence and GGM. Specifically, we observe that along the $\lambda=0$ line, $\Delta^C_\mu \approx 0$ in Case 3, while  $\Delta^C_\lambda >0$ for Case 2 (See Fig.~\ref{fig:N6dxyozzcmztzzggm} and \ref{fig:N6oxydzzcmztzzggm}). 



\begin{figure}
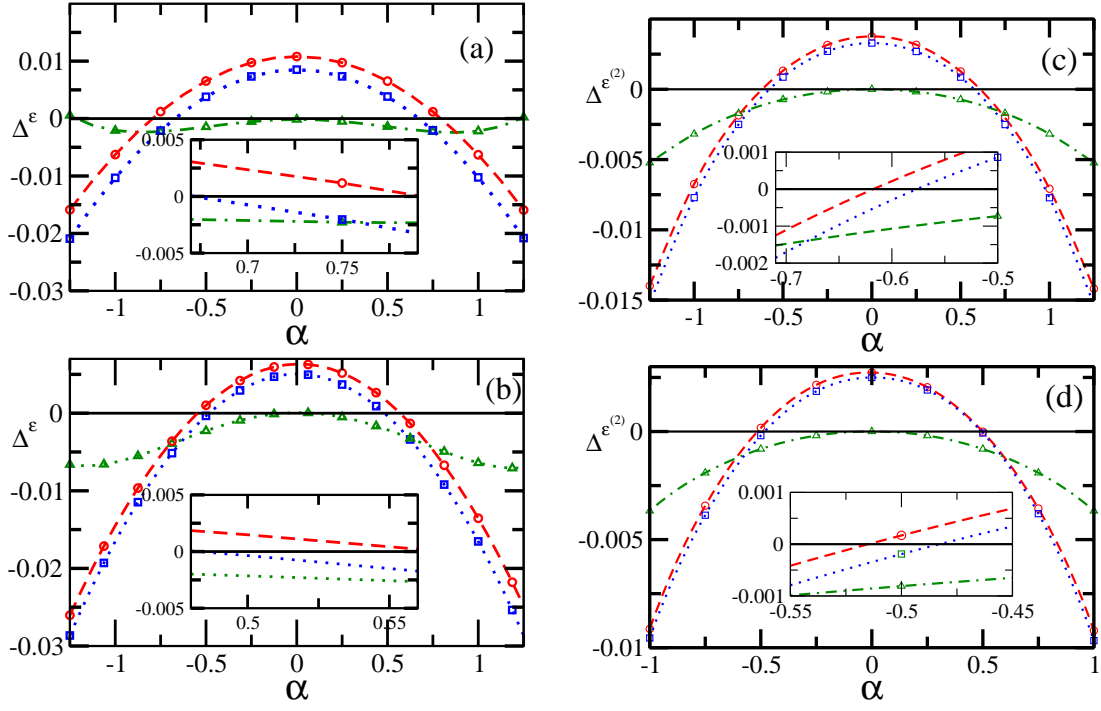

\vspace*{0 cm}
\includegraphics[angle=0,width=70mm]{Figure-4-Sen.eps}\hspace{+0.5cm}
\includegraphics[angle=0,width=70mm]{Figure-6-Sen.eps}
\vspace*{0 cm}
\caption{ Constructive interference. The panels exhibit plots with the enhancement score of GGM, denoted as $\Delta^{{\cal E}}$, as the ordinate, and the system parameter, $\alpha$, as the abscissa for Heisenberg spin glasses with (a) $N = 6$, (b) $N = 8$, and the enhancement score of approximate GGM, denoted as $\Delta^{{\cal E}^{(2)}}$ with (c) $N = 12$, and (d) $N = 16$, where $N$ is the number of quantum spin-1/2 particles in the system. In panels (a) and (b), red circles connected with dashed lines represent the cases when disorder is present in both the couplings (planar as well as azimuthal), and for these cases, $\alpha$ represents $\lrang{\lambda}$ and $\Delta^{{\cal E}}$ represents $\Delta^{{\cal E}}_{\lambda,\mu}$. We choose $\lrang{\delta} = -0.9$. The blue squares connected with dotted lines are for the cases when disorder is present only in the planar coupling. In these cases, $\alpha$ represents $\lrang{\lambda}$, $\Delta^{{\cal E}}$ represents $\Delta^{{\cal E}}_{\lambda}$, and $\delta$ is fixed at $-0.9$. The green triangles connected with dash-dotted lines represent the cases when disorder is present only in the azimuthal coupling. In these cases, $\alpha$ represents $\lambda$, $\Delta^{{\cal E}}$ represents $\Delta^{{\cal E}}_{\mu}$, and $\lrang{\delta}=-0.9$. The black solid lines, parallel to the horizontal axes are drawn to separate the positive and negative regions of the enhancement score of GGM. The insets show blow-ups of the regions with constructive interference. The depiction of the cases for $\Delta^{{\cal E}^{(2)}}$ in panels (c) and (d) are analogous to the cases for  $\Delta^{{\cal E}}$ shown in the panels (a) and (b).
%
All other descriptions are the same as in Fig.~\ref{fig:N6dxydzzcmztzzggm}. The insets show blow-ups of the regions with constructive interference. For all the plots, we have chosen $\gamma = 0.7$ and $h = 0.8$. All quantities plotted here are dimensionless. 
}
\label{fig:extggmN}
\end{figure}
\section{Constructive interference}
The Heisenberg Hamiltonian that we study involve planar and azimuthal interaction strengths, which may both be Gaussian distributed quenched disordered variables. Our calculations show that irrespective of whether they are individually or jointly present, a spectrum of measurable quantities show disorder-induced enhancement, instead of getting diminished in the presence of defects. At this juncture, we ask a more radical question: {\it Does there exist any observable which gets enhanced in the joint presence of the disorders while it deteriorates when the randomness is applied individually in either of the couplings, in the Heisenberg spin glass models?}  Mathematically, we are looking for the following conditions to be satisfied simultaneously
by an observable ${\cal Q}$:

\begin{eqnarray}
\Delta^{{\cal Q}}_{\lambda,\mu}&=&|{\cal Q}_{av}(\lrang{\lambda},\lrang{\mu})|-|{\cal Q}(\lrang{\lambda},\lrang{\mu})|>0, \label{eq:enhcondtn1}\\
{\textrm{while \hspace{1.5em}}}  & \nonumber \\ 
\Delta^{{\cal Q}}_{\lambda}&=&|{\cal Q}_{av}(\lrang{\lambda})|-|{\cal Q}(\lrang{\lambda})|<0,\label{eq:enhcondtn2}\\
\Delta^{{\cal Q}}_{\mu}&=&|{\cal Q}_{av}(\lrang{\mu})|-|{\cal Q}(\lrang{\mu})|<0.
\label{eq:enhcondtn3}
\end{eqnarray}
Here, $\mu = \lrang{\mu}$ in Eq.~(\ref{eq:enhcondtn2}) and ${\cal Q}_{av}(\lrang{\lambda})$ corresponds to the planar spin glass. Similarly, $\lambda=\lrang{\lambda}$ in Eq.~(\ref{eq:enhcondtn3}) and ${\cal Q}_{av}(\lrang{\mu})$  corresponds to the azimuthal spin glass. Any observable satisfying the above set of equations would imply that the competing random interactions can interfere constructively for the quantity ${\cal Q}$.  We refer to this phenomenon as the ``constructive interference of ${\cal Q}$''.
We have extensively investigated the above equations for various finite number of spins in the Heisenberg Hamiltonian ranging from $N = 6$ to $N=20$, and scanned over significant ranges of the system parameters. While smaller system sizes are handled by exact diagonalization, the relatively larger ones are investigated by employing the density matrix renormalization group techniques \cite{dmrg1}. The investigations help us to identify certain parameter ranges, 
where the system exhibits the phenomenon of the constructive interference  in the genuine multipartite quantum correlation as quantified by the generalized geometric measre (GGM), denoted by \(\mathcal{E}\)\cite{GGM}. Interestingly, no other observable, considered here, shows such a behavior. This includes (single-site) magnetizations, two-site classical correlators, and two-site  entanglement. 

 Let us first discuss the results corresponding to the Heisenberg spin glass models, consisting of a comparatively small number of spins. The behavior of quantum correlations, in this case can be examined  in experimentally realizable systems like  ion traps, photons, etc \cite{ion-trap-1,ion-trap-2,photon,opt-lattices,supercqubit-1,supercqubit-2,supercqubit-3}, and hence can be verified and observed in the laboratories. All the results presented in this case are obtained by performing exact diagonalization of the Hamiltonians. Fig.~\ref{fig:extggmN} shows the enhancement scores for the GGM for different system sizes, and for different blending of disordered couplings. The anisotropy and external magnetic field strength are again chosen as $0.7$ and $0.8$ respectively. The coupling strength, $\delta$, in the azimuthal direction, is $-0.9$ for the case when it is ordered, while in the cases when the azimuthal couplings, $\delta_{ij}$, are disordered, they are chosen with mean $-0.9$ and unit standard deviation. 

Whenever any of the curves, in the panels of Fig.~\ref{fig:extggmN}, is positive, the corresponding disordered system has higher value of GGM as compared to that of the ordered system. 
It is clear from Figs.~\ref{fig:extggmN}(a-b) that for GGM, $\Delta^{{\cal E}}_{\lambda}$, $\Delta^{{\cal E}}_{\mu}$, as well as $\Delta^{{\cal E}}_{\lambda,\mu}$ are positive, showing disorder-induced enhancement although in different parameters ranges with the system sizes $N=6$ and $N=8$ respectively. Note that there exists other parameter ranges than those exhibited in the panels of Fig.~\ref{fig:extggmN}(a)-(b) where such phenomenon occurs. The choice of the parameters and parameter ranges in Figs.~\ref{fig:extggmN} are for the following specific purpose. Near the two values of $\alpha$ $(=\lrang{\lambda},$ here), where the curves of $\Delta^{{\cal E}}_{\lambda,\mu}$ crosses the horizontal axes, one obtains regions where disorder-induced enhancement for GGM is exhibited with the introduction of both planar and azimuthal disorders, while the same is absent with the inclusion of just any one of these disorders. We call these as ``Venus regions''.  For example, for $N=6$  (see Fig.~\ref{fig:extggmN}(a)), the two distinct ranges of $\alpha$ (which represents either $\lrang{\lambda}$ or $\lambda$), in which the constructive interference can be observed are $[-0.78, -0.67]$ and $[0.68, 0.78]$. In these regions, the enhancement score, $\Delta ^{\mathcal{E}}_{\lambda,\mu}$ (red circles connected by dashed line), is positive while the other two enhancement scores, viz., the $\Delta ^{\mathcal{E}}_{\lambda}$ (blue squares connected by dotted line) and $\Delta ^{\mathcal{E}}_{\mu}$ (green triangles connected by dot-dashed line), remain negative.
Note that with increasing number of particles, the Venus regions, i.e., the windows of $\alpha$ demonstrating constructive interference moves towards $\alpha=0$. Interestingly, no such phenomenon is found in other quantities considered in this paper, viz., magnetization, two-point correlators, and bipartite entanglement (see Figs.~3(a) and 3(b)).  It is worth mentioning here that the Venus regions do not surface without an external magnetic field.
In fact, depending on $N$, there exists a critical magnetic field strength, $h_c$, only beyond which the constructive interference can  be observed. Below $h_c$, the $\Delta ^{\mathcal{E}}_{\lambda}$ lies above the $\Delta ^{\mathcal{E}}_{\lambda,\mu}$ and the $\Delta ^{\mathcal{E}}_{\mu}$. As the external magnetic field is increased beyond $h_c$, the curve corresponding to the $\Delta^{\mathcal{E}}_{\lambda,\mu}$ goes above that of  $\Delta^{\mathcal{E}}_{\lambda}$ and $\Delta^{\mathcal{E}}_{\mu}$, resulting in the emerging of the Venus regions.  We find that the $h_c$ is approximately $0.75$ for $N=8$.

\begin{figure}
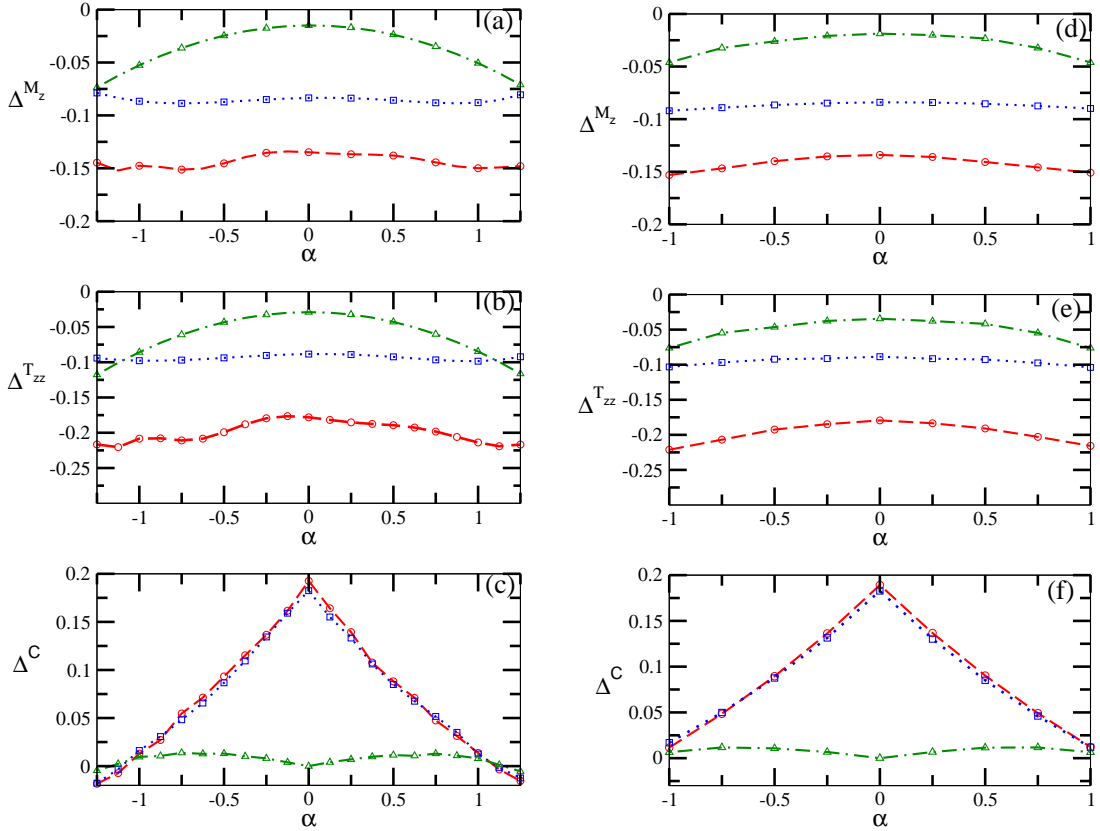

\vspace*{+0.2cm}
\includegraphics[angle=0,width=70mm]{Figure-5-Sen.eps}\hspace{+0.5cm}
\includegraphics[angle=0,width=70mm]{Figure-7-Sen.eps}
\vspace*{-0.2cm}
\caption{ The enhancement scores for (a) magnetization, $M_z$, (b) $T_{zz}$-correlator, and (c) concurrence, $C$, against $\alpha$ for  systems with $N=6$. Same quantities are again shown in (d), (e), and (f) for $N=20$ calculated using  DMRG method discussed in \ref{appendix_exact_dmrg}. All other descriptions are the same as given in Fig.~\ref{fig:extggmN}.}
\label{fig:allQN6}
\end{figure}


 It is natural to ask if the results presented above also holds for the disordered Heisenberg systems with larger number of spins. However, exact computation of the quenched averaged GGM, in systems with large number of parties, is hindered, due to the following three key reasons: 
$(i)$ An exponential growth of the Hilbert space with increasing number of parties essentially prohibits one from performing exact diagonalization of the Hamiltonian.
 in the systems with large number of spins. 
$(ii)$ For obtaining the desired accuracy, as one tries to obtain convergence in the quenched averaging of physical quantities, it typically requires a large number (approximately $5\times 10^3$ to $8\times 10^3$) of random realizations, unless the quantities are self-averaging,
which is not the case for all genuine multipartite observables. 
$(iii)$ Determination of  multipartite entanglement, as quantified by the GGM, requires all possible bipartite splits, and the number of bipartitions in an $N$-party system is $\sum_{r = {1}}^{N/2} {n \choose r}$, which  increases substantially with increasing $N$. For example, the number of bipartitions required to evaluate the GGM for the $N=8$ system is 162, whereas it grows to over half a million for the system involving just $20$ parties.

The difficulty in computing GGM can partly be curbed by choosing selective bipartitions instead of considering all possible bipartitions. We therefore introduce ${\cal E}^{(2)}$ as a measure of multipartite entanglement, defined as
${\cal E}^{(2)} = 1-\max\left \{ \{\eta_i^2\},\{\eta_{i,i+1}^2\}\big|_{i=1,\ldots,N} \right \}$,
where $\eta_i$ and $\eta_{i,i+1}$ are the maximum Schmidt coefficients of the single- and nearest-neighbor two-body reduced density matrices respectively. We call it the ``approximate GGM''. Although ${\cal E}^{(2)}$ may not be a {\it genuine} multiparty entanglement measure, it does quantify multiparty entanglement and it is certainly an entanglement monotone.

 For the Heisenberg spin models with and without disorder, 
we evaluate the approximate GGM and  plot the enhancement scores in Fig.~\ref{fig:extggmN}(c-d). 
Here we consider the spin systems with sizes $N=12$ (Fig.~\ref{fig:extggmN}(c)), and $N=16$ (Fig.~\ref{fig:extggmN}(d)). The symbols are kept consistent with Figs.~\ref{fig:extggmN}(a-b).  It can clearly be noticed that the $\Delta^{\mathcal{E}^{(2)}}$ again identify two distinct ranges of the parameter $\alpha$ (on the negative and positive sides of $\alpha$), where the Venus regions materialize. We also find that the conclusions drawn from the ${\cal E}^{(2)}$ are consistent with the physics discussed by studying the system with smaller sizes. For example, the windows of $\alpha$ exhibiting constructive interference shifts towards $\alpha=0$  with increasing number of spins. We have carried out analysis upto $N=20$ and confirm the existance of a non-zero region, where constructive interference occurs. It is to be noted that the shrinking observed, of the Venus regions, could be due to the modification of the multiparty entanglement measure, since we have already observed that no two-party or single-site observables exhibit the constructive interference. 
It is plausible that in the presence of both the disorders, multipartite entanglement will exhibit a Venus region even in the thermodynamical limit. 

{\color{black} At this point, it will be interesting to look for generalized features in the disordered system that may be specific to quantum phases of the corresponding ordered system. Particularly, it is interesting to identify the phase of the corresponding ordered system where the phenomenon of constructive interference appears due to the application of disorder. The homogeneous spin-1/2 quantum XYZ model exhibits a rich phase diagram, consisting of the antiferromagnetic (AFM) phase, magnetized or polarized phase (PP), spin-flop (SF) phase, and floating phase (FP) (see Ref. \cite{Jonas} and references therein). While we find that the phenomenon of disorder-induced enhancement is not phase-specific, that of constructive interference appears in the so-called SF phase, which is characterized by N{\' e}el ordering in the y-direction for small values of $\mu/\lambda$.} 

{\color{black}  While we have used the GGM as a measure of multiparty entanglement, there do exist other measures. 
It would be interesting to learn if the phenomena of disorder-induced enhancement and constructive interference are observed in other measures of multiparty entanglement. 
For this purpose, in addition to GGM, we consider another measure of multiparty entanglement, viz. the tangle (denoted as $\tau$, and also known as the monogamy score for squared concurrence) \cite{CKW}. We examine its behavior as a function of $\alpha$ for a selective set of parameters for which we have reported the occurrence of disorder-induced enhancement and constructive interference for GGM in this work. The results indicate that disorder-induced enhancement of multiparty entanglement is a potentially generic feature. On the other hand, constructive interference does not occur in the parameter region that supported the same for GGM.}

We additionally investigate the magnetization, two-point correlations in the $zz$ direction and the concurrence \cite{concurrence} in this regime.  Figures 3(c) and 3(d) show their behavior for $N=12$ and $N=16$, respectively. As already mentioned, we find no constructive interference phenomena in any of these quantities.  Note that the results for these larger systems are obtained by using DMRG technique, which consider the spin chains to be open (see Methods for details). For $N$-site systems, the bipartite classical and quantum correlations are calculated for the $(N/2,N/2+1)$ pairs, so that boundary effects are minimized. The magnetization is calculated for the $N/2^{th}$ site. As it can be readily understood, for systems with periodic boundary condition, which we have used for smaller systems, one is free to select any site or pair of neighboring sites for calculating a given single- or two-site observable, as then the quenched averaged quantities are site independent. However when we use DMRG, one cannot rule out the boundary effects on the physical quantities{\color{black}, particularly the local ones including one- and two-body quantities,} even for large systems. {\color{black} Global observables, like genuine multisite entanglement, are not expected to change with the introduction of a single interaction (required for going from open to periodic boundary condition). Nevertheless, we find that a satisfactory description of the phenomena under study viz., disorder-induced enhancement and constructive interference, is possible, even for the local observables, with open DMRG, provided the system size is not too small and the measurement of the observables on either fringe is excluded. This has been confirmed by a simple test, where we examine the quantitative values of disorder-averaged magnetization at the adjacent sites located at the center of the chain $((N/2)^{th}$ and $(N/2+1)^{th}$ sites), and the  bipartite quantities corresponding to the adjacent bipartite sub-systems at the center, constituted of the $(N/2 - 1, N/2)$ and $(N/2, N/2 + 1)$ site pairs. A close quantitative agreement of the quantities corresponding to both the cases implies that the qualitative features of the local quantities remain unfazed by the end effects.} 

\section{Conclusion}
\label{sec:conclusion}
In summary, we have studied  the quantum Heisenberg spin system in one-dimension with random coupling interactions. We have examined the behavior of the magnetization, classical as well as the two-party quantum correlations, and  multipartite entanglement for the ground states. The relevant results are presented for various system sizes ranging from five to twenty quantum spin-1/2 particles. While the small systems were dealt  by exact numerical diagonalization, we adopt the density matrix renormalization technique to investigate comparatively larger spin systems. In the presence of impurities in the couplings, there exists different parameter regions for the different observables which show enhancement due to disorder -- also known as the disorder-induced enhancement phenomenon. The physical quantities like magnetization, classical correlators, bipartite and multipartite entanglement always find a range of parameters in which they  increase with the introduction of disorder. Perhaps more radically, our studies uncover the novel phenomenon of constructive interference, where we observe that the parameters of the system can be tuned in such a way that disorder-induced order appears due to simultaneous presence of randomness in two different couplings, while it is absent when disorder is present individually in either of the couplings.

The constructive interference, which is caused due to the interplay between competing random coupling strengths in different directions, appears only in the multipartite entanglement, and is absent in bipartite as well as single-site physical quantities considered, exhibiting 
the significance of multiparty observables in cooperative physical phenomena. It would 
be interesting to understand the connection of these phenomena with localization-delocalization transition at finite temperature \cite{sesher-sedin}. 

\appendix


\section{Methods:} 
\label{appendix_exact_dmrg}
\subsection{Exact diagonalization and DMRG technique.}
The smaller spin systems can be simulated by directly  diagonalizing the Hamiltonian matrix. However, the same is not true for larger spin systems due to the exponential  growth of the Hilbert space, due to which  the computations are either highly expensive or computationally hard. In order to perform numerical simulations for systems with a larger number of spins, we adopt the finite-size density matrix renormalization group method \cite{dmrg1} with the open boundary condition, which is an iterative numerical approach for obtaining highly accurate low energy physics of  quantum many-body systems. We choose to work with open boundary conditions as it is well known that the accuracy drops significantly for closed boundary conditions. For $N$-site systems, the bipartite classical and quantum correlations are calculated for the $(N/2,N/2+1)$ pairs, so that boundary effects are minimized. In the DMRG approach, starting from a portion of the system, known as system block, the system size is enlarged  step by step until the desired system size is reached. At each step, the Hilbert space is truncated by retaining only those Schmidt basis vectors which corresponds to the $m$ highest Schmidt coefficients $\lambda_i$, $i=1,2, \cdots,m$, of the block reduced density matrix. 
The value of the physical quantities for the disordered spin chain are achieved by performing several sweeps of the finite system DMRG \cite{dmrg1}.

\begin{figure}
\centering
\label{test}
\vspace*{+0.2cm}
\includegraphics[angle=0,width=65mm]{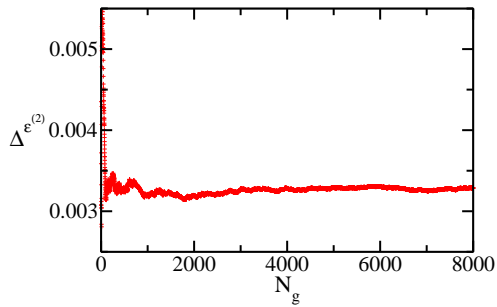}
\vspace*{0.2cm}
\caption{(Color online.) Illustration of convergence during quenched averaging. $\Delta^{{\cal E}^{(2)}}$ as a function of the number of random realizations, $N_g$, for the case when the Hamiltonian belongs to case 2,  with $\langle \lambda  \rangle=0$ for $N=12$. All quantities plotted are dimensionless.}
\label{fig:convergence}
\end{figure}
 \subsection{Quenched averaging}
 \label{appendix_Quenched_averaging}
In the disordered models, that we have considered, the physical parameters of the system  are ``quenched", i.e., the time scales in which the dynamics of the system takes place is much shorter in comparison to the time in which the disordered system parameters equilibrate.  It implies that during the {\color{black}observation-time} of the system,  a particular realization of the random disorder parameters remains frozen. The physically relevant values {\color{black}determining the equilibrium statistical properties } of the system observables (physical quantities) are, therefore, their quenched averaged values, where we first compute the value of the physical quantity of interest for a given disorder configuration of the system and subsequently perform the averaging over the probability distribution of the disorder. 

{\color {black} The term ``quenched" in context of randomness can be seen as having been derived from the  process of sudden cooling (i.e., quench) of a liquid to obtain a glass, which is devoid of any long-range structural order, with atoms trapped in random positions, resulting in ``quenched" disordered competing interactions in couplings.  The glass is however heading towards its structural equilibration extremely slowly and will take a very large time to do so. Therefore, for the observation time for a given observable, the configuration is understood to be fixed.}

{\color{black} In this work, the random parameters are chosen from independent and identically distributed Gaussian distributions with a given mean ($\mu$) and standard deviation ($\sigma$). This, in principle, allows the random couplings to vary from -$\infty$ to +$\infty$. However, as the random parameters are chosen from a Gaussian distribution, the coupling coefficients, in reality, are roughly confined within $\mu - 3 \sigma$ to $\mu + 3 \sigma$. The probability of a randomly chosen coupling appearing from extreme fringes is understandably small. The cases coming from extreme fringes have a rather negligible contribution.   As a result, the quantities smoothly converge towards fixed values with increasing number of random realizations.} We test convergence
of the quenched averaged quantities with the increase in number of Gaussian distributed random configurations, $N_g$. We find that it typically requires a few thousand
of random numbers to reach the desired convergence. Fig.~(\ref{fig:convergence})  shows an example of convergence for $\Delta^{{\cal E}^{(2)}}$, where the convergence is up to the third decimal point.

\section{Magnetization and  classical correlators}
 \label{appendix_quantities}
The physical quantities that we have studied are single-site observables like magnetization, two-site observables like classical correlators, and bipartite as well as multipartite quantum correlation measures. For the Heisenberg Hamiltonian, that we consider, in both ordered as well as disordered cases, the $x$ and $y$ components of the magnetization of the ground state vanish, while the $z$ component of the magnetization, 
$M_z^{i}=\textrm{Tr}(\sigma^z_i\rho^A_{i})$,  of the single-site reduced density matrix ($\rho^{A}_{i}$) at the $i^{\textrm{th}}$ site of the ground state is in general non-vanishing. 
In the disordered case, one has to further perform a quenched averaging over the relevant variables to obtain the physically meaningful quenched averaged magnetization. The classical correlators between the $i^{\textrm{th}}$ and $j^{\textrm{th}}$ sites are defined as $T_{\alpha\beta}^{ij} = \textrm{Tr}(\sigma^{\alpha}_{i}\otimes\sigma^{\beta}_{j}\rho^{AB}_{ij})$ with $\alpha,\beta=x,y,z$ and with $\rho^{AB}_{ij}$ being the bipartite density matrix obtained from  the ground state. It can be shown that the off-diagonal correlators of the ground state vanish in both ordered and disordered cases \cite{vanish-off-corr-1, vanish-off-corr-2, vanish-off-corr-3}.

\textbf{Concurrence.}  In case of bipartite entanglement, we first found the $N$-party ground state of a given Hamiltonian,  and then  we trace out all the $(N-2)$ parties except two nearest-neighbour sites and subsequently considered the entanglement of that two-party state. In this work, we have used the concurrence \cite{concurrence} as the bipartite entanglement measure. The concurrence, $C_{AB}$, of the nearest-neighbour bipartite state, $\rho^{AB}$, is defined as
$C_{AB} = \textrm{max}\{0,\lambda_{1}-\lambda_{2}-\lambda_{3}-\lambda_{4}\}$,
where $\lambda_{i}\, (i= 1,2,3,4)$ are the square roots of the eigenvalues of the matrix $\rho^{AB}\widetilde{\rho^{AB}}$, in decreasing order, and $\widetilde{\rho^{AB}} = (\sigma^{y}_{A}\otimes\sigma^{y}_{B}){\rho^{*}}^{AB} (\sigma^{y}_{A}\otimes\sigma^{y}_{B}$), where complex conjugation is  with respect to the computational basis. Concurrence is a
monotonically increasing function of the entanglement of formation, which in turn is defined as the average entanglement required to create the bipartite quantum state, modulo certain 
additivity problems \cite{concurrence}.

\textbf{Genuine multiparty measurement (GGM).} As a measure of genuine multiparty entanglement, we have employed the GGM \cite{GGM} (cf. \cite{GM}). In case of quantum systems composed of more than two subsystems, the quantification of entanglement is much more involved in comparison to the bipartite case. This is because in a multipartite scenario, there are qualitatively  different  kinds of entangled states like bi-separable, tri-separable, etc., and  there are also genuine multipartite entangled states. For pure multiparty states, genuine multiparty entangled states are defined as those which are not product across any bipartition. In order to quantify genuine multiparty entanglement, we use the GGM which is based on the distance between the $N$-party   pure state,  and an $N$-party pure state which is not genuinely multiparty entangled. More specifically, the GGM for an $N$-party pure quantum state, $\arrowvert \psi_{N} \rangle$, is given by
 ${\cal E}(|\psi_{N}\rangle)=1-\max|\langle \phi_{N}| \psi_{N}\rangle|^{2}$,
where the maximization is taken over all $N$-party pure quantum states, $|\phi_{N}\rangle$, that are not genuinely multiparty entangled. It is possible to evaluate the maximization analytically for an arbitrary state $ |\psi_{N}\rangle$ and is given by
 ${\cal E}(|\psi_{N}\rangle)=1-\max\{\eta_{{\cal A}:{\cal B}}^{2}|{\cal A} \cup {\cal B}=\{1,\ldots,N\}, {\cal A}\cap {\cal B}=\emptyset\}$,
where $\eta_{{\cal A}:{\cal B}}$ is the maximal Schmidt coefficient of $|\psi_{N}\rangle$ in the bipartite split ${\cal A} : {\cal B}$. The GGM gives us a ``distance'' of the multiparty state under consideration from the set of states which are not genuinely multiparty entangled.

\begin{flushleft}
{\bf Acknowledgments}
\end{flushleft}
We acknowledge the useful discussions with Luigi Amico during the meeting on Quantum Information Processing and Applications (QIPA-2013) held at the Harish-Chandra Research Institute (HRI), India. RP acknowledges support from the Department of Science and Technology, Government of India, in the form of an INSPIRE faculty scheme at HRI. We acknowledge computations performed at the cluster computing facility in HRI. This work has been developed by using the DMRG code released within the ``POWDER with Power'' project (www.dmrg.it).


\begin{flushleft}
{\bf References}
\end{flushleft}

\end{document}